# Multi-faceted Sensory Substitution for Curb Alerting: A Pilot Investigation in Persons with Blindness and Low Vision

Ligao Ruan, Giles Hamilton-Fletcher, Mahya Beheshti, Todd E Hudson,

Maurizio Porfiri, JR Rizzo

**Abstract**: Curbs -- the edge of a raised sidewalk at the point where it meets a street -- crucial in urban environments where they help delineate safe pedestrian zones, from dangerous vehicular lanes. However, curbs themselves are significant navigation hazards, particularly for people who are blind or have low vision (pBLV). The challenges faced by pBLV in detecting and properly orientating themselves for these abrupt elevation changes can lead to falls and serious injuries. Despite recent advancements in assistive technologies, the detection and early warning of curbs remains a largely unsolved challenge. This paper aims to tackle this gap by introducing a novel, multi-faceted sensory substitution approach hosted on a smart wearable; the platform leverages an RGB camera and an embedded system to capture and segment curbs in real time and provide early warning and orientation information. The system utilizes YOLO (You Only Look Once) v8 segmentation model, trained on our custom curb dataset for the camera input. The output of the system consists of adaptive auditory beeps, abstract sonification, and speech, conveying information about the relative distance and orientation of curbs. Through human-subjects experimentation, we demonstrate the effectiveness of the system as compared to the white cane. Results show that our system can provide advanced warning through a larger safety window than the cane, while offering nearly identical curb orientation information.

Introduction

Curbs are critical and ubiquitous features in urban landscapes, marking the boundary between a raised sidewalk or road median and a street or other roadway [1]. For people who are blind or have low vision (pBLV), these urban features can become hazardous, impeding safe and efficient navigation. Without the visual cues that sighted individuals rely on to detect elevation changes, pBLV may not perceive curbs until it is too late, often leaving them with a limited window to react safely and/or orientate themselves correctly [2]. This lack of awareness may lead to trips and falls, perhaps most concerning for the pBLV who are elderly [3][4]. At present, more than 25% of adults aged 71 years and older in the United States are visually impaired [5].

To address these challenges, decades of development have resulted in bevies of assistive technologies [6]. Recent advances increasingly utilize artificial intelligence [7], particularly computer vision [8], to dynamically interpret complex urban environments. These assistive applications can detect obstacles and recognize key infrastructural elements, thereby boosting the confidence of visually impaired individuals and expanding their capacity for independent urban mobility [9]. A few of these approaches [10][11] can detect the curb or find travelable paths from curb cues with limited accuracy or sensitivity, but the lack of distance and orientation information in curb feedback limits effectiveness and, ultimately, adoption.

We propose a novel system in which a custom-trained YOLOv8 instance segmentation model [12] accurately extracts curb information from simple RGB camera input. The feedback of our system integrates multi-faceted, audio-based sensory substitution, which translates the visual data into auditory signals that pBLV users can interpret [13]. This method utilizes auditory beeping [14], sonification [15] and speech to convey information about detected curbs' relative distance and orientation. Beeping sounds are used to indicate proximity, scaling in intensity as the user gets closer to the curb. Sonification transforms orientation data into non-speech audio cues, which for sensory substitution, involves the conversion of visuospatial information into variations of frequency, amplitude, and panning over time to produce rich, visually meaningful soundscapes [16]. Speech output is offered as an alternative through text-to-speech technology to deliver descriptive audio messages [17], focused on the description of a user's orientation relative to the curb. Integrating these auditory feedback mechanisms aims to provide pBLV with multi-faceted, visual-to-auditory sensory substitution, enhancing spatial awareness of the curb and reducing the risk of curb-caused injury.

The architecture of the proposed system is articulated into three primary steps. The first step recognizes curbs from diverse visual inputs through a custom-trained instance segmentation model. The second step extracts the distance information and map the segmentation result to the alert zone which is divided into three zones: far, medium and near. The final step translates distance and the curb's orientation through multi-faceted auditory prompting. Curb distance estimations are communicated through short beeps with adapting frequency, duration, interpulse interval, reverb, and loudness. The orientation information of the curb is transmitted as sonification or speech outputs. Sonification utilizes the vOICe algorithm [18] to represent the curb orientation and adds a spatial panning effect, and speech output uses text-to-speech technology to provide text-based information about orientation to the user. To assess the system, we conducted human-subject experiments with pBLV to measure the effectiveness of the multi-faceted sensory substitution in guiding mobility; we evaluated both the safety window (distance to curb at notification) and orientation accuracy (curb angle detected) as dependent variables.

**Methods**

*Curb detection and segmentation*

Our system uses the YOLOv8 instance segmentation (YOLOv8-Seg) model [19] for curb detection for its real-time, single-shot detection capabilities [20] that are ideal for applications requiring immediate feedback, such as obstacle warning devices. YOLOv8-Seg model extends the YOLOv8 object detection framework by integrating instance segmentation capabilities, utilizing the CSPDarknet53 as its backbone [21]. This backbone, a convolutional neural network based on the DarkNet-53 design, employs a CSPNet strategy that bifurcates the feature map of the base layer into two segments, subsequently merging them using a cross-stage hierarchical structure [22]. The neck processes these features, often using mechanisms like feature pyramid networks (FPN) or path aggregation networks (PAN) to refine features at different scales. Additionally, YOLOv8-Seg introduces a novel C2f module, replacing the traditional YOLO neck architecture, which streamlines processing and bolsters performance and efficiency. The head of the YOLOv8-Seg model comprises two parts: the detection head, which predicts bounding boxes and class probabilities, and the segmentation head, which predicts segmentation masks by upscaling feature maps and applying pixel-wise classification. Despite these advancements, the model maintains a structure similar to YOLOv8, featuring five detection modules and a prediction layer, thereby preserving the system's robust detection capabilities. Post-processing steps, such as non-maximum suppression (NMS) and thresholding, are then applied to refine the model's outputs.

In the initial phase of the study, we created a custom dataset containing approximately 3,000 images of urban curbs in New York City (NYC) from a pedestrian's perspective, each manually labeled with a polygonal outline of the curb (shown in Figure 1) for training the YOLOv8-Seg model. The YOLOv8-Seg model can output bounding boxes and masks of detected curbs, but only the masks are used to convert information about the location and contour of curbs in our system.

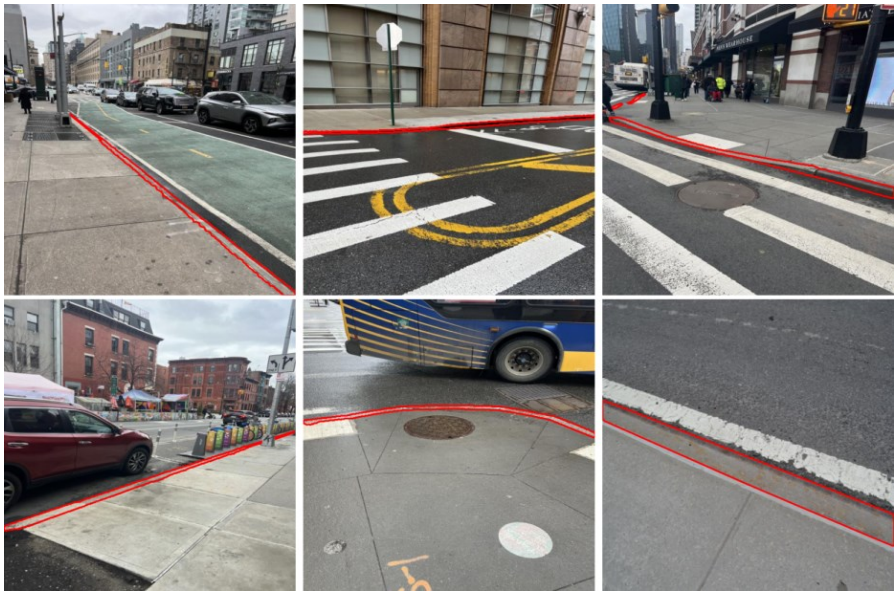

Figure 1. Examples from our curb dataset. These images are captured from a pedestrian perspective and include various scenes and angles. Each curb is labeled with polygons to train the instance segmentation model.

Before training the YOLOv8-Seg model, we implemented a series of image augmentation techniques to enhance the model's robustness and accuracy under varied environmental conditions and mitigating overfitting [23][24]. These techniques, including horizontal and vertical flips, introduce the model to diverse orientations, mimicking the variability encountered in real-world settings. Shear adjustments of up to ±10° in both directions help the model adapt to angular distortions typical in camera-based imaging. Brightness variations of up to ±20% are achieved by adjusting the RGB values by factors ranging from 0.8 to 1.2, preparing the model for different lighting conditions, while Gaussian blurs of up to 2 pixels and noise affecting up to 0.77% of pixels simulate minor imperfections in image capture that occur due to focus issues or environmental interference [25].

*Alert zones design*

In the second step, we devised a method for communicating the relative distance to the curb using predefined proximity 'windows', thereby defining the adaptive auditory prompts. Our system employs a region delineation method similar to that used in automotive reversing cameras, using three nested layers of a minor sector to represent near, medium, and far distances, respectively. The center of the sector, which is out of the frame, represents the user's location towards the bottom of the image, as shown in Figure 2. The detected distances to the curb are grouped into these proximity zones. Here, we effectively use the vertical position in the image as a proxy for distance of floor-based objects. Given basic assumptions of the camera (height and tilt angle) and assuming a flat ground surface, we can accurately approximate curb distances through their position in the image frame. Here, the camera was placed at a height of approximately 135 cm from the user and tilted downward by 30 degrees, with the far region covering a distance from 146 cm to 257 cm in front of the user, the medium region covering a distance from 90 cm to 146 cm, and the near region covering a distance up to 90 cm in front of the user as shown in Figure 3.

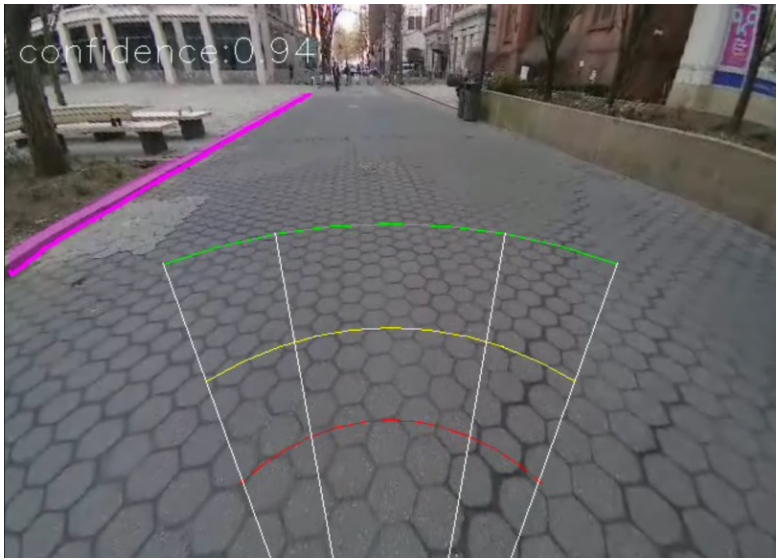

Figure 2. Alert zone visualization. Red, yellow, and green curves indicate the outer extent of the near, medium, and far zones, respectively. Each color zone refers to the space between the line of the corresponding color and the line of the next color closer to the bottom of the image. The far zone refers to the part between the top green line and the lower yellow line, the medium zone refers to the part between the yellow line and the red line, and the near zone refers to the part between the red line and the bottom of the image. The lower contour line of the detected curb is shown in pink. Here only the closest curb is visualized.

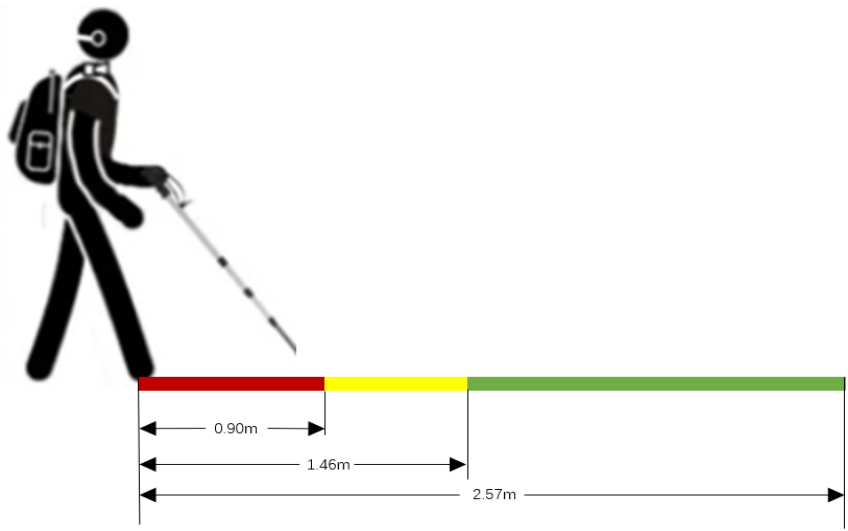

Figure 3. Side view of the alert zone. Red, yellow, and green colors indicate near, medium, and far alert zones respectively.

If more than one pixel of the mask enters the proximity alert zone, the mask will be considered within that zone (see Figure 4) and the corresponding alert will be triggered. The distance to the curb is calculated by finding the minimum pixel distance from the mask to the center of the sector. This distance is then used to determine which alert zones the curb enters. The average slope of the lower contour of the mask is calculated to determine the orientation of the curb. When more than one curb is detected, the curb belonging to the pixel closest to the center of the sector will be chosen.

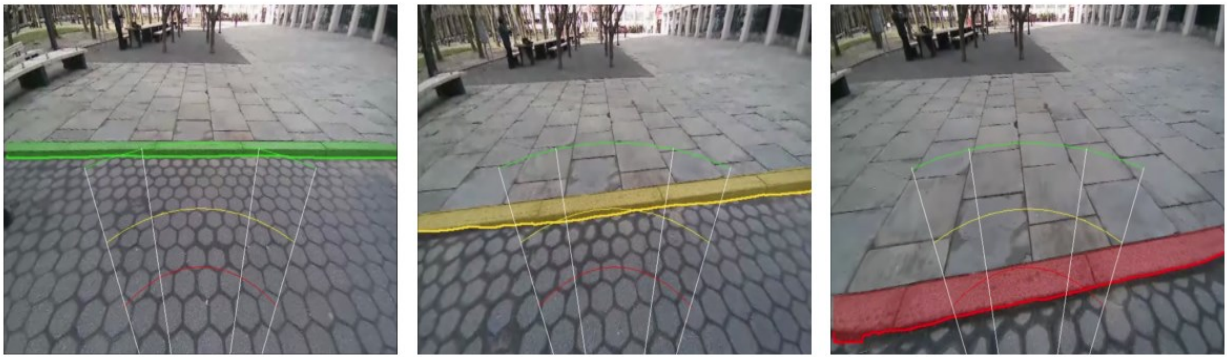

Figure 4. Visual representation of the three alert zones (far, medium, and near) designed for curb detection. The left image shows a curb entering the far zone, highlighted in green. The middle image shows a curb entering the medium zone, highlighted in yellow. The right image shows a curb entering the near zone, highlighted in red.

*Audio alert design*

*Proximity alert*

The auditory cues for curb proximity are designed to change as the user approaches the curb.

Each alert zone (far, medium, and near) is associated with distinct parameters: frequency, beep duration, interpulse interval (IPI) of beep, reverberance, and loudness. Frequency, measured in hertz (Hz), indicates the pitch of the beep, which increases as the user gets closer to the curb. Beep duration specifies the length of time each beep lasts, decreasing as the user gets closer to the curb. The IPI of beeps, measured in seconds, decreases as the user approaches the curb, resulting in more frequent beeps. Reverberance, expressed as a percentage, indicates the echo effect applied to the sound and decreases as the user gets closer to the curb. Loudness expressed as a percentage related to the device volume setting increases as the user approaches the curb. See Table 1 for specific parameter settings.

The far zone is divided into four sub-levels, while the medium zone has three sub-levels. To maintain clear distinctions between the main alert zones, we apply smaller steps for sub-levels within each zone and larger steps between zones. This way, the gradual shifts within a zone don't blur the bigger steps between zones, keeping the alert progression intuitive for the user. For example, in the far zone, the frequency changes by 15 Hz and the IPI changes by 0.2 seconds between sublevels, whereas in the medium zone, the frequency changes by 50 Hz and the IPI changes by 0.15 seconds between sublevels. As a result, as users approach the curb, multiple auditory dimensions change, with larger perceptual changes occurring as the curb line crosses from one alert zone into another. This means that users can either pay attention to a specific auditory characteristic, or to overall levels of auditory change and use this to discriminate their distance to the curb [26]. These auditory changes allow users to gauge their distance from the curb through specific auditory characteristics, providing clear and distinguishable signals that enhance their ability to respond promptly and maintain safety.

| Alert Zone | Distance (cm) | Frequency (Hz) | Duration (s) | IPI (seconds) | Reverberance (%) | Loudness (%) |
|---|---|---|---|---|---|---|
| Far | 257 - 230 | 205 | 0.07 | 1.5 | 40 | 80 |
| | 229 - 202 | 220 | | 1.3 | | |
| | 201 - 174 | 235 | | 1.1 | | |
| | 173 - 146 | 250 | | 0.9 | | |
| Medium | 145 - 123 | 300 | 0.06 | 0.8 | 30 | 100 |
| | 123 - 107 | 350 | | 0.65 | | |
| | 106 - 90 | 400 | | 0.5 | | |
| Near | Up to 90 | 500 | 0.05 | 0.4 - 0.2 | 20 | 120 |

Table 1: Auditory beep parameters of curb proximity alerts.

*Orientation alert*

The purpose of the orientation alert is to assist the user with orientating themselves correctly to safely step up or down the curb or turn and avoid it altogether. In the design for indicating the orientation of the curb, two audio feedback options were provided: sonification and speech, with sonification refreshed every 3 seconds and speech every 4 seconds, including their respective duration times. These methods utilize the average slope calculated above, rounded to the nearest five-degree interval to simplify feedback design. The orientation alert is designed to operate independently of the proximity alert, ensuring that it does not interfere with the ongoing beeping pattern.

*Speech output.* To generate the speech output, our system utilizes the Mimic3 text-to-speech engine, which generates clear and natural-sounding vocal audio. Mimic3 is based on the VITS model [27], a non-autoregressive model designed to directly predict speech waveform from input text. Such a model allows for rapid and efficient speech synthesis. The feedback design takes as its baseline the angle at which the user would need to turn to be perpendicular to the curb. For example, if the calculated gradient is positive 30 degrees, the audio feedback needs to indicate that the user would be perpendicular to the curb if they turned 30 degrees to the left. Therefore, the system will issue a voice prompt "30 left".

*Sonification.* Our sonification method is inspired by the vOICe system, a vision-to-audio sensory substitution system introduced in 1992 by Peter Meijer. The system captures images via a camera and converts them into soundscapes, which are delivered to the user through headphones, conveying one image frame / soundscape per second [18]. Each soundscape involves a horizontal scan of the visual scene from left to right over a one second sweep of time without panning. An image frame is sampled once per second and converted to grayscale, white pixels are transformed into frequency of sounds representing the vertical position within the image and loudness indicating brightness.  As a result, users experience a continuous series of auditory "snapshots" of their environment.

To utilize the vOICe output, we created a series of images that represent the orientation of the curbs (see Figure.5). Each image shows the curb at specific angles, starting at 0 degrees (horizontal line) and increasing in 5-degree increments up to 175 degrees. The curbs are presented as dashed white lines on a black background to enhance visual and auditory contrast and clarity when sonified. Dashed lines provide more salient auditory changes than solid lines and align with previous work [28]. The angle of the curb is gradually increased in a clockwise direction, with each image slightly rotating the dotted line from the previous image to indicate a gradual change in orientation. These images were processed using the vOICe application, converting each image into 0.8-second audio clips. The conversion translates the direction of the dashed line into sound patterns of varying pitch and volume corresponding to the angle, e.g., in a 145-degree image, the line crosses diagonally from the lower left to the upper right corner, producing a sound where the frequency rises over time. In the vOICe system, verticality is represented by pitch, and laterality is represented by time.

*Spatial feedback.* The original audio files generated by the vOICe system do not have the panning effect, so all sounds were presented without spatial differentiation in the left-right auditory field. In our system, these audio files and the proximity beeps are spatialized to provide spatial auditory cues to the users. The panning effect in the audio representing the curb orientation is achieved by adjusting the gain applied to the left and right audio channels based on the pan position, which varies linearly from a starting point to an endpoint from left to right along the lower contour of the curb mask. The pan position is calculated using equation (1), normalizing the relative horizontal position of the sound source to a range between -1 (full left) and 1 (full right).

$$pan_{orientation} = (\frac{x_{pixel}}{frame\ width}) \times 2 - 1 \qquad (1)$$

where $x_{pixel}$ is the pixels' current horizontal position along the curb mask's lower contour, and $frame\ width$ is the total width of the frame.

The left and right channel gains are then adjusted according to this pan position to create a stereo effect. The gains are computed using equations (2) and (3). The spatialized stereo signal is finally obtained by applying these normalized gains to the mono audio signal, converting it to a stereo format.

$$Left\ gain_v = \frac{(1-pan_{orientation})^2}{Norm\ factor} \quad (2)$$

$$Right\ gain_v = \frac{(1+pan_{orientation})^2}{Norm\ factor} \quad (3)$$

$$Norm\ factor = \sqrt{(1-pan_{orientation})^2 + (1+pan_{orientation})^2} \quad (4)$$

The spatial effect of the proximity beeps is achieved by a similar method. The gains of the left and right audio channels are determined by calculating the pan position based on the pixel position that is closest to the center of the alert zone sector. The gains for the left and right audio channels are calculated as:

$$Left\ gain_p = 1 - \frac{x_{closest\ pixel}}{frame\ width} \quad (5)$$

$$Right\ gain_p = \frac{x_{closest\ pixel}}{frame\ width} \quad (6)$$

where $x_{closest\ pixel}$ is the horizontal position of the closest pixel of the curb mask, and $frame\ width$ is the total width of the frame.

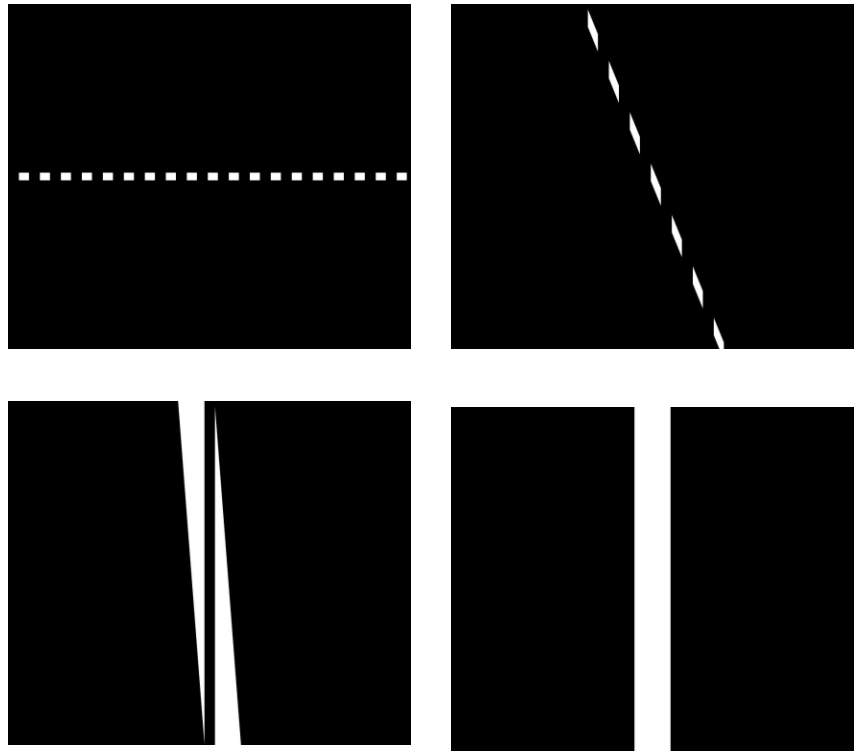

Figure 5. Examples from the orientation representation images. We artificially thickened the line width when the white line segments approached ninety degrees to enhance the audio clips produced by vOICe, making them more pronounced.

*Participant demographics*

Ten participants with blindness/low vision were recruited for this study to test the effectiveness of our system. Three participants were congenitally blind, while seven developed significant vision impairments later in life (acquired blindness). The mean age of the participants was 51.3 ±14.88 years, and nine received formal training in using the white cane. Four of these participants currently use a guide dog as their primary mobility aid. Basic demographic information about the participants is presented in Table 2. Participants were also asked about their difficulty in judging curbs, steps, stairs, or uneven pavement due to their vision impairment. The responses were categorized as follows: four participants reported no difficulty, two participants found it slightly difficult, three participants found it moderately difficult, and one participant found it very difficult.

| Participant | Age | Visual Ability | Current Mobility Aid |
|---|---|---|---|
| 1 | 32 | No Light Perception | Family Member |
| 2 | 59 | No Light Perception | Cane |
| 3 | 55 | No Light Perception | Guide Dog |
| 4 | 71 | No Light Perception | Family Member |
| 5 | 33 | Minimal Light Perception | Cane |
| 6 | 36 | Minimal Light Perception | Guide Dog |
| 7 | 42 | Minimal Light Perception | Cane |
| 8 | 47 | Minimal Light Perception | Cane |
| 9 | 63 | Minimal Light Perception | Cane |
| 10 | 75 | Minimal Light Perception | Cane |

Table 2. Participant demographics.

*Experiment design*

To access the effectiveness of this system, we recruited participants with blindness or low vision to use a white cane either with or without assistance from our system in experiments that require curb discrimination in a natural outdoor environment at St. Vartan Park in NYC. Participants were tasked with walking toward the curb from a starting location 3 meters away from the curb with the initial approach angle (θ) at 0˚, 30 ˚, and 60 ˚ (see Figure 6). Upon detecting the curb, participants needed to stop and reorient their bodies based on the feedback provided by the navigation aid so that they could walk parallel to the curb to mimic the avoidance of a tripping hazard. Before the experiment, a 5-minute tutorial was provided to each subject to familiarize them with the system and the audio feedback.

In the control group, using the traditional white cane, participants stopped and adjusted their orientation to be parallel to the curb once they touched the curb with the cane. In the experimental groups, participants stopped once they heard the medium level of beeps and then reoriented their bodies based on the audio feedback to align themselves parallel to the curb. The safety window (d), which is the distance between the stopping point and the curb, was measured and recorded. The angle difference (θ) between the participant's final orientation and the curb orientation was measured and recorded. These steps were repeated for each combination of navigation conditions (cane alone, beeps with sonification, and beeps with speech output) and starting angles (0˚, 30˚, and 60˚). The trial order was varied for each new subject to minimize any potential order effects. The collected data were used to evaluate the effectiveness of the navigation aids.

Participants were equipped with a customized navigation system comprising several components. A backpack was utilized to house the primary processing unit, an NVIDIA Jetson Orin NX, along with a Krisdonia 25000mAh power bank to ensure sufficient power supply throughout the experiment. Mounted on the strap of the backpack was an IMX258 13MP USB camera, positioned to capture the participant's forward-facing view. Audio feedback was provided through SHOKZ OpenComm2 Open-Ear Bone Conduction Headphones, which allowed participants to receive auditory instructions while maintaining environmental awareness.

In addition, all participants wore the same occluding glasses during the experiment to ensure a consistent level of visual impairment across all trials.

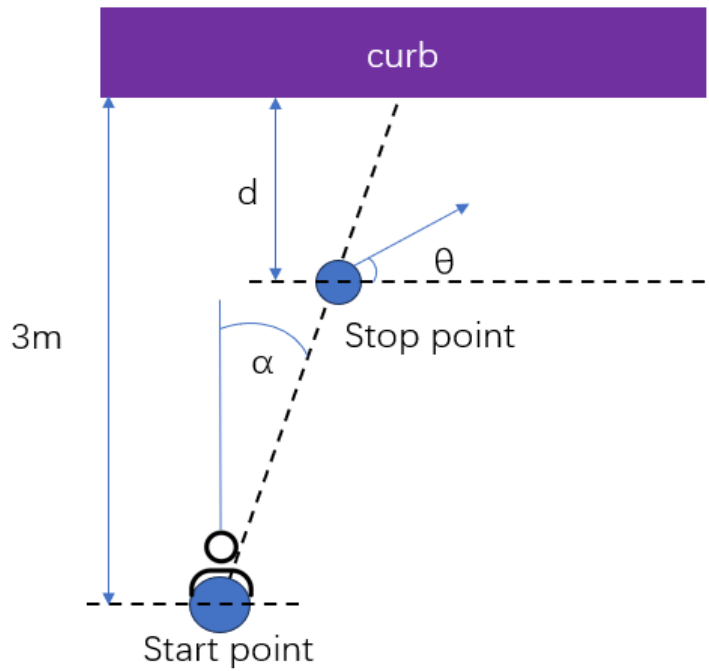

Figure 6. Procedure of the experiment for assessing the effectiveness of navigation aids. Participants started to walk toward the curb 3 meters away with the initial approach angle (θ) at 0°, 30 °, and 60 °, then stopped based on the feedback from the navigation aid and reoriented their bodies to be parallel to the curb. The safety window (d) and angle difference (θ) were measured and recorded.

*Statistical analysis*

Data was analyzed using repeated-measures ANOVA with three navigation conditions (cane alone, beeps with sonification, and beeps with speech output), three starting angles (0°,30°, and

60°) at a level of significance of 0.05, and post hoc analyses were performed using Tukey's HSD [29].

**Results**

*Curb Segmentation*

After fully trained with our curb dataset for 70 epochs as shown in Figure 7, the curb segmentation model performs well in detecting and segmenting curbs. In the validation phase of our YOLOv8-Seg model, we conducted a thorough evaluation across the test dataset comprising 563 images with 925 curb instances. The model demonstrated good performance metrics for mask predictions, achieving a precision of 0.84 and a recall of 0.74. Pixel accuracy is calculated as 0.93 with the equation (7). The intersection over Union (IoU) is calculated as 0.83 with the equation (8). The processing speed was efficient, with an average of 1.0ms for preprocessing (resizing or normalizing frames), 40.8ms for inference, and 2.9ms for postprocessing involving non-maximum suppression per image with Jetson Orin Nx.

$$Pixel\ Accuracy = \frac{p_{correct}}{p_{total}} \qquad (7)$$

where $p_{correct}$ is the number of pixels correctly identified as 'curb' and represents the total number of 'curb' pixels in the ground truth. It provides a measure of how many pixels in the segmentation output exactly match their corresponding labels in the ground truth.

$$IoU = \frac{|Predicted\ Mask \cap Ground\ Truth\ Mask|}{|Predicted\ Mask \cup Ground\ Truth\ Mask|} \qquad (8)$$

where the *Predicted Mask* is the output generated by the segmentation model, and *Ground Truth Mask* is a binary image created from the annotation. A higher IoU indicates a better performance of the segmentation model, as it shows a greater alignment between the predicted and actual regions of interest.

Mean Average Precision (mAP) is another common evaluation metric used in object detection tasks to measure the accuracy of the model's predictions. Our model achieved a mAP50 of 0.789 and a mAP50-95 of 0.55. mAP50 refers to the mean average precision calculated at an Intersection over Union (IoU) threshold of 0.50. This means that a predicted bounding box is considered correct if it overlaps with the ground truth bounding box by at least 50%. mAP50-95, on the other hand, is the average mAP calculated across multiple IoU thresholds ranging from 0.50 to 0.95, providing a more comprehensive evaluation of the model's performance across different levels of localization precision.

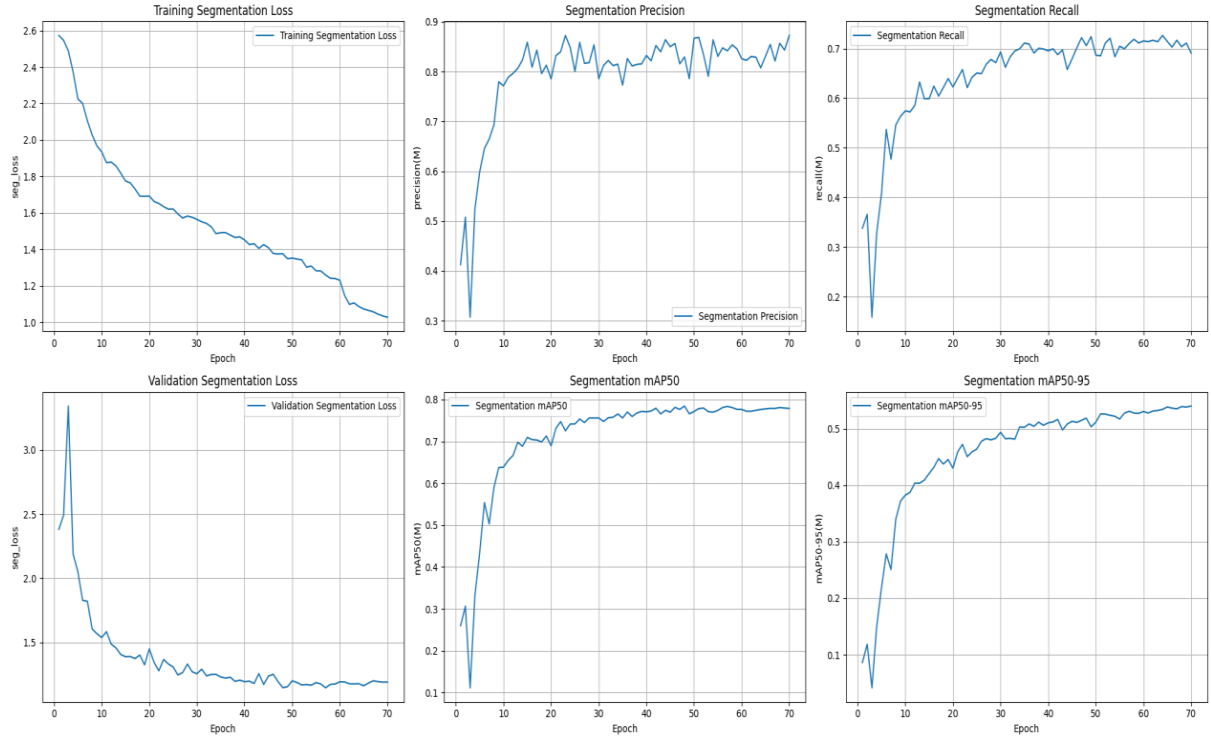

Figure 7. Training and validation performance metrics for YOLOv8-Seg Model. This figure illustrates various training and validation metrics for the YOLOv8-Seg model over 70 epochs. The top row depicts the training segmentation loss, segmentation precision, and segmentation recall. The training segmentation loss shows a decreasing trend, indicating the model's learning progress. Segmentation precision and recall metrics increase and stabilize over the epochs, reflecting improved model performance. The bottom row presents the validation segmentation loss, segmentation mAP50, and segmentation mAP50-95. The validation segmentation loss decreases, while both segmentation mAP50 and mAP50-95 metrics improve, suggesting that the model generalizes well to unseen data.

*Human-Subjects Data analysis*

*Safety window*

The boxplot in Figure 8 illustrates the safety window distributions across different navigation conditions and starting angles. The x-axis represents the combination of navigation conditions and starting angles, while the y-axis represents the safety window measurements. Higher y-values indicate enhanced performance, as a result of larger safety windows. The error bars, shown as whiskers, indicate the spread of the data, with the range covering the minimum and maximum values within 1.5 times the interquartile range (IQR) from the quartiles. The "beeps with sonification" condition show the largest spread, indicating higher variability in stopping distances. The "cane alone" condition has the smallest IQR, suggesting more consistent stopping distances. Outliers are present in both the beeps with sonification and cane alone conditions, reflecting individual trials with significantly different stopping distances.

Repeated measures ANOVA was conducted to examine the effects of navigation conditions and starting angles on the safety window revealing significant results. The main effect of the navigation condition on the safety window was statistically significant ($F_{(2, 18)} = 17.57$, $p = 0.0001$), indicating that navigation condition significantly influenced the stopping distance from the curb. The main effect of the starting angle did not reach statistical significance ($F_{(2, 18)} = 2.30$, $p = 0.1294$), suggesting that the starting angle did not significantly impact the safety window, indicating that all approaches were resilient to approaching the curb from different angles. Additionally, the interaction effect between navigation condition and starting angle was not statistically significant ($F_{(4, 36)} = 1.35$, $p = 0.2711$), indicating that starting angle did not have unique effects on specific navigation conditions for the resultant safety window.

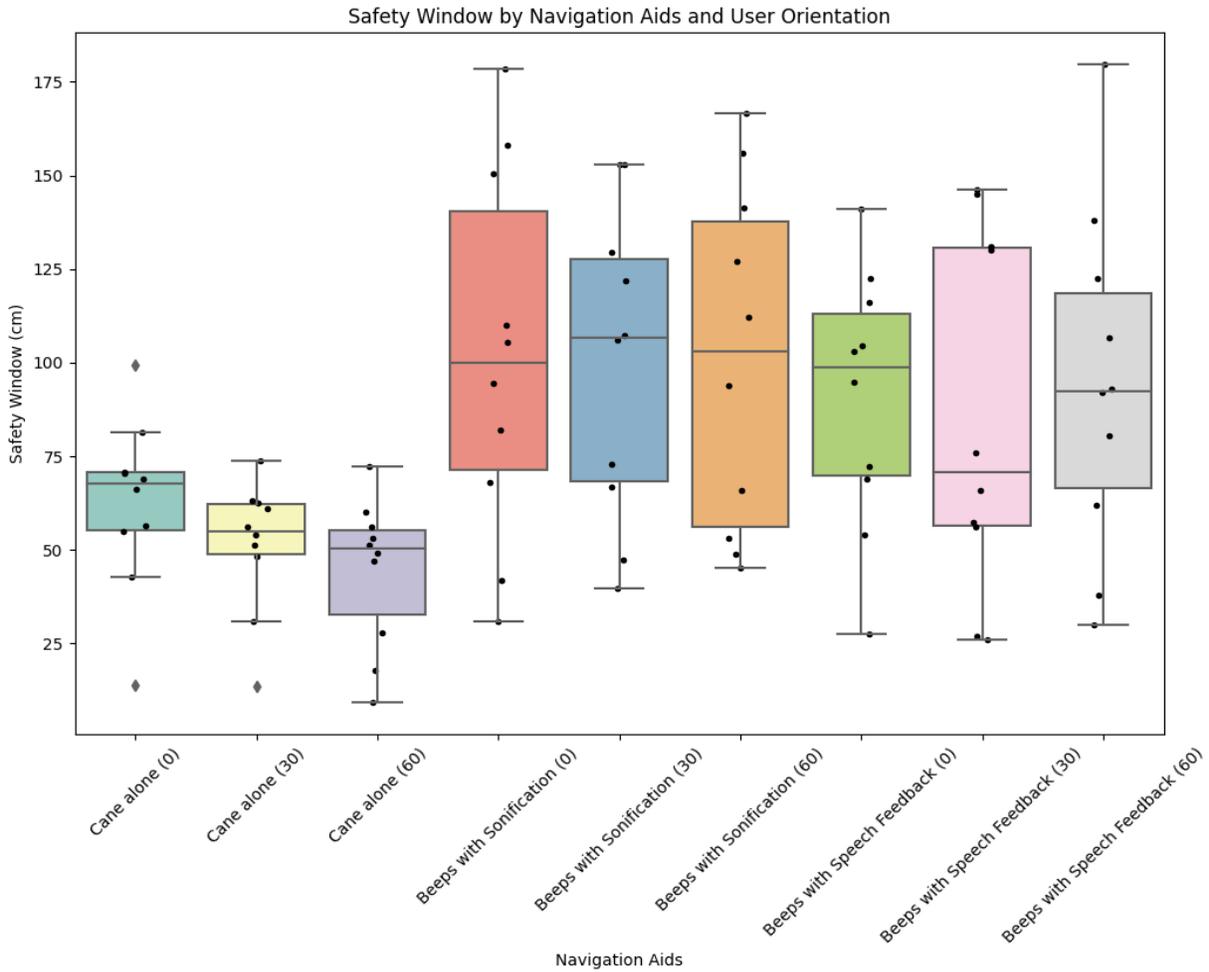

Figure 8. Safety window by navigation conditions and starting angle. The x-axis represents the different combinations of navigation aids and starting angles (0, 30, and 60 degrees). The y-axis represents the safety window measurements. Higher y values indicate better performance. The error bars, depicted as whiskers extending from each box, cover the minimum and maximum values within 1.5 times the IQR. Outlier is indicated by rhombus.

Following the significant ANOVA results for the safety window, a post hoc analysis using Tukey's Honest Significant Difference (HSD) test was conducted to identify specific differences between the navigation conditions. The results indicated that the mean difference in stopping distance between the "beeps with sonification" condition and the "cane alone" condition was significant ($p < 0.0001$), with a mean difference of -48.11, suggesting that the "beeps with sonification" condition significantly improved the stopping distance compared to the "cane alone" condition. The comparison between the "beeps with speech output" condition and the "cane alone" condition also revealed a significant difference ($p = 0.0005$), with a mean difference of -37.42, indicating that the "beeps with speech output" condition significantly improved the stopping distance. The difference between the "beeps with sonification" and "beeps with speech output" conditions was not significant ($p = 0.5046$), with a mean difference of -10.69. These results highlight that both auditory navigation aids ("beeps with Sonification" and "beeps with speech output") significantly enhance the safety window compared to using a traditional white cane alone, while there is no significant difference between the two auditory aids.

*Orientation relative to the curb*

This measurement represents the angle at which the participants are oriented relative to the curb when they stop (shown in Figure 9). An angle of zero degrees indicates that the participant is perfectly parallel to the curb. In this orientation, if the participant were to walk forward, they would walk alongside the curb rather than towards it. Positive or negative angles indicate deviations from this perfect parallel orientation, where the participant is either angled towards or away from the curb, respectively.

The ANOVA conducted to examine the effects of navigation condition and starting angles on the orientation relative to the curb revealed the following results. The main effect of the navigation condition did not have a statistically significant effect ($F(2, 18) = 0.3550, p = 0.7060$) on participants' selection of angles to position themselves to be parallel with the curb. The main effect of the starting angle approached significance ($F(2, 18) = 3.3376, p = 0.0585$), suggesting a potential influence of the starting angle on the angle difference; however, it did not reach the threshold for statistical significance. Additionally, the interaction effect between navigation condition and starting angle was not statistically significant, indicating that the interaction effect of navigation condition and starting angle did not significantly impact the angle difference from the curb ($F(4, 36) = 0.8533$, $p = 0.5011$). This result suggests that the navigation aids have a similar effect to the traditional white cane regarding the angle at which participants position themselves relative to the curb.

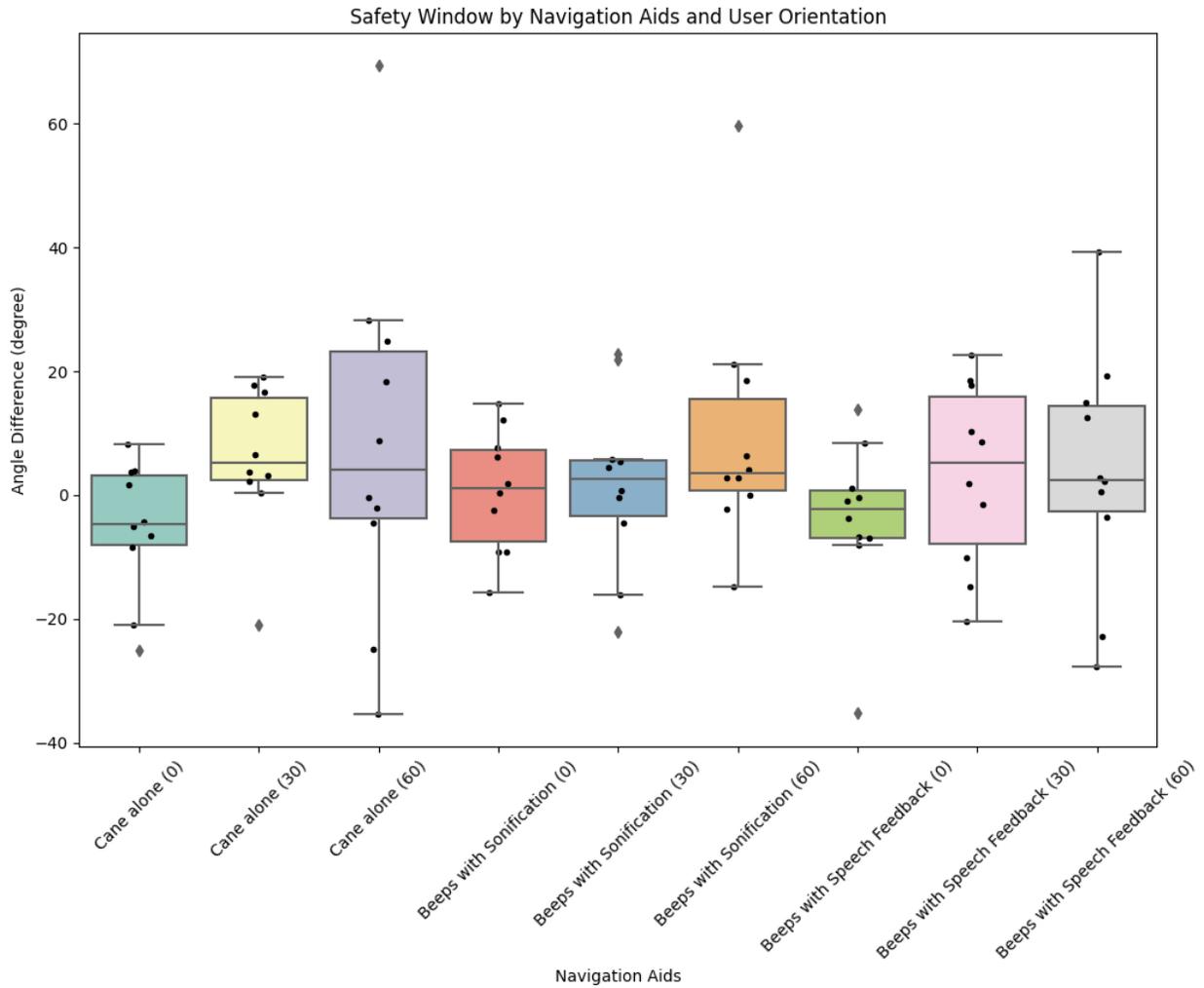

Figure 9. Angle difference from the curb by navigation conditions and starting angle. The x-axis represents the different combinations of navigation aids and starting angles (0, 30, and 60 degrees). The y-axis represents the angle difference (measured in degrees) between the user's orientation and the curb orientation, with values closer to zero indicating better performance. The error bars, depicted as whiskers extending from each box, cover the minimum and maximum values within 1.5 times the IQR. Outlier is indicated by rhombus.

**Discussion**

Results from our experiments indicate that the proposed system can effectively detect curbs and provide helpful auditory feedback to users, enhancing their spatial awareness and safety when approaching the curb. The segmentation model trained with our dataset achieved good performance with precision of 0.84, recall of 0.74, pixel accuracy of 0.93 and IoU of 0.83. The experiment with pBLVs also supports the notion that our system is able to provide a timely proximity and orientation alerting of the curb through multi-faceted sensory substitution.

*Curb Detection through Visual Analytics*

Segmentation-based approaches have shown promise but most are still in the process of being fully developed and optimized. Furthermore, only a limited number of studies have applied this method to assistive systems for pBLVs [11].

This approach typically involves performing a semantic segmentation task to roughly identify line-shaped objects such as lanes and curbs, followed by the application of post-processing algorithms to refine the results. Kailun Yang et al. proposed using pixel-wise semantic segmentation to aid visually impaired individuals in navigating curbs and improving traversal awareness along pathways. Their model achieved 42.92% accuracy for the curb class [10]. Yichao Cai et al. developed a semantic segmentation model named FCN-HRNet-S, utilizing an

HRNet-like encoder and an FCN decoder to achieve smooth and continuous curb segmentation in the field of autonomous driving. Their models achieved a high level of accuracy, but this was only demonstrated with images specifically of curbs following cleaning [30]. Diwei Sheng et al. investigated the efficacy of curb segmentation in foundational models. They discovered that several state-of-the-art models struggle with high false-positive rates (up to 95%) and exhibit poor performance in distinguishing curbs from similar objects or non-curb areas, such as sidewalks [31]. These studies have shown that while curb segmentation is promising, it still requires improvements to be an effective curb detection method in the field of navigation aids for pBLVs.

The accuracy of curb detection specifically for pedestrian use needs to be further improved. One of the reasons for this limitation is that the training of these models needs more support with standardized datasets containing images of pedestrian viewpoints. While there are large-scale datasets like BDD100K [32], which offers 100,000 labeled images, its focus is primarily on the driver's view, which doesn't align well with pedestrian needs. ADE20K [33] includes diverse scene annotations but only provides 331 curb labels, which is insufficient for training robust curb detection models. Similarly, datasets like Mapillary Vistas [34] and Cityscapes [35], though comprehensive in urban environments, lack enough detailed curb annotations relevant to pedestrian viewpoints.

In addition, the detection performance of the model can be affected by lighting conditions, while weather factors such as snow, fog, and rain can also reduce accuracy. These adverse conditions obscure salient curb features, introduce noise, and cause more interference. Our custom dataset will be improved in the future to address these issues.

*Hardware Embodiment and Integration*

During our tests, we observed that the camera angle can deviate due to accidental collisions, leading to variations in camera placement. These variations in the angle can result in inconsistencies in the detection and therefore feedback provided by the system when encountering the same curb at the same distance. In this study, we determined the design of the alert zone based on a pre-determined height and angle of the webcam relative to the ground. Unlike surveillance systems or automobile autopilot systems, the webcam is mounted on the shoulder straps of a backpack worn by the user, making its position more susceptible to changes due to gait-induced motion and accidental collision during mobility. This variability can lead to inconsistencies in calculating the relative distance or orientation of the curb. To avoid the necessity of setting the camera at a specific height and angle, future advancements could incorporate 3D sensors, depth estimation models, or building an auto-calibration system. 3D sensors like LiDAR [36] or stereo cameras [37] provide precise depth information regardless of camera orientation. Depth estimation can infer depth from 2D images using deep learning models, adapting to various conditions [38]. Auto-calibrating systems can also dynamically adjust camera parameters using horizon detection, compensating for user movement and ensuring consistent curb detection. Integrating these technologies may enhance the system's reliability and flexibility may make the technology more adaptable to real-world scenarios.

Another limitation in our system's current design is the assumption of a flat ground surface. The curb distance is estimated based on the assumption that the user is navigating on flat terrain, which may not always be the case in real-world environments. Sloped sidewalks, uneven paths, or other irregularities can distort distance calculations and lead to inaccurate alerts. Addressing this issue requires exploring more adaptable solutions, such as integrating depth sensors or enhanced depth estimation algorithms that can dynamically account for variations in ground elevation.

*Multi-faceted Audio-based Sensory Substitution*

Vision-to-audio sensory substitution systems collect visuospatial information from the surrounding environment and then translate this information directly into auditory representations. As such, end users can learn to associate specific auditory feedback with specific visual features, allowing pBLV to perceive otherwise inaccessible sensory details. While this approach preserves a high degree of visual detail in the resultant soundscape, complicated visual images become difficult to interpret as changes in luminance/loudness do not necessarily correlate with information that is pertinent for the user. As a result, in urban environments, the sonification results produced by these sensory substitutions can be too complicated for users to extract key information, and thus require a considerable amount of practice by the pBLV to be mastered for functional tasks [39].

The form of sensory substitution can be intuitive for simple shapes or patterns [40] but requires many months of training to accurately interpret raw visual images of more complicated imagery [41]. However, simplifying complex visual scenes into only the most relevant and meaningful contrasts may be more usable [42][43]. Furthermore, sonification design that reflects theories of human hearing can further improve user performance [44]. Some attempts to make this style of sonification more practical for end users include the sonification of color, depth maps, and thermal images, as seen in the Depth-vOICe, Synesthesia, and SoundSight [45]. Here the contrasts produced by these forms of visuospatial information better align with specific tasks such as object identification, obstacle avoidance, and localization of people, animals, and hazards. As a result, soundscapes may be simpler (e.g., only nearby object silhouettes get sonified), easier to interpret (e.g. loudness = object proximity) and more actionable. Rather than sonifying 'snapshots', the SoundSight App provides real-time updates from changes in the input image to the resultant soundscape. This allows users to immediately detect changes in the environment and better orientate themselves to navigate through obstacles. However, while these approaches can be helpful in some scenarios, the selection of specific object categories in the environment may require additional processing such as through object detection and/or segmentation methods.

In our study, we designed a multi-faceted system in the auditory domain that incorporates two concurrent modes of feedback, both beeping and sonification, or beeping and speech output, making the information more immediately usable for the user. While the optimal sonification methods for conveying changes in frequency, beep duration, IPI of beep, reverb, and loudness are yet to be determined, each of these parameters can be fine-tuned to enhance the clarity and intuitiveness of the auditory feedback in our future work. Additionally, specific design requirements may be necessary to accommodate different forms of visual impairment. For instance, individuals with varying degrees of hearing loss or auditory processing abilities might require personalized adjustments to the sonification parameters. Tailoring the system to address these diverse needs can ensure that the assistive technology is effective and accessible to a broader range of users.

Another critical aspect to consider is distinguishing between curb-up and curb-down scenarios. This could be achieved by analyzing the relative height and position of curbs in the segmented output. For instance, when a curb-up is detected, the system could use a high-low tone sequence to indicate an upward step, while a curb-down scenario could be signaled with a low-high tone sequence to indicate a downward step. These distinct auditory cues would help users discern the type of curb they are approaching, enhancing the overall navigational assistance the system provides.

**Conclusion**

In this study, we developed and evaluated a multi-faceted, audio-based sensory substitution system designed to help pBLVs detect curbs and receive timely alerts about their relative distance and orientation. Our findings indicate that the system provides a larger safety window compared to a cane-only scenario, demonstrating its potential to enhance user safety. The system did not show significant differences in alignment accuracy compared to the traditional white cane; this suggests that it performs similarly in helping users maintain proper orientation with the curb.

In the future, efforts will focus on improving curb detection accuracy through the development of larger curb datasets and more optimal technological embodiments including the integration of novel 3D sensors and depth estimation AI approaches. Additionally, addressing the variability in camera placement with auto-calibration methods and stabilizing equipment will further enhance the system's reliability and ease of use. Lastly, exploring creative sonification methods and tailoring the system to meet the specific needs of users with different forms of visual impairment will also enhance system usability.

Acknowledgment

The work was supported by the National Science Foundation [ITE2345139, DUE2129076, CNS1952180]; National Institutes of Health [R33EY033689]. The content is solely the authors' responsibility and does not necessarily represent the official views of the National Institutes of Health and the National Science Foundation.

NYU, John-Ross Rizzo, and Todd E. Hudson have financial interests in related intellectual property. NYU owns a patent licensed to Tactile Navigation Tools. NYU, John-Ross Rizzo, and Todd E. Hudson are equity holders and advisors of said company.**Reference**

1. Momotaz, H., Rahman, M., Karim, M., Iqbal, A., Zhuge, Y., Ma, X., Levett, P. (2022). A Review of Current Design and Construction Practice for Road Kerbs and a Sustainability Analysis. *Sustainability*, *14*(3), 1230. https://doi.org/10.3390/su14031230

2. Legood, R., Scuffham, P., Cryer, C. (2002). Are we blind to injuries in the visually impaired? A review of the literature. *Injury Prevention*, *8*(2), 155-160. https://doi.org/10.1136/ip.8.2.155

3. Harwood, R. (2001). Visual problems and falls. *Age and Ageing*, *30*(suppl 4), 13-18. https://doi.org/10.1093/ageing/30.suppl_4.13

4. Cullinan, T. Visual Disability and Blindness. In *Visual Disability in the Elderly* (pp. 1-13). Routledge. https://doi.org/10.4324/9781032698243-1